\documentclass[12pt]{article}
\usepackage{amsfonts}
\usepackage{amssymb}
\usepackage{amssymb}
\usepackage{graphicx}
\usepackage{amsmath}
\usepackage{cite}

\begin{document}

\title{The Affine Hidden Symmetry and Integrability of Type IIB Superstring
in $AdS_{5} \times S^{5}$ }
\author{Bo-Yu Hou$^{a}$ \thanks{%
Email:byhou@nwu.edu.cn}, Dan-Tao Peng$^{a b}$ \thanks{%
Email:dtpeng@nwu.edu.cn or dtpeng@ustc.edu.cn}, Chuan-Hua Xiong$^{a}$
\thanks{%
Email:chxiong@phy.nwu.edu.cn}, Rui-Hong Yue$^{a}$ \thanks{%
Email:rhyue@nwu.edu.cn} \\
$^{a}$Institute of Modern Physics, Northwest University,\\
Xi'an, 710069, P. R. China \\
$^{b}$Interdisciplinary Center for Theoretical Study, \\
University of Science and Technology of China, \\
Hefei, Anhui, 230026, P. R. China}
\date{}
\maketitle

\begin{abstract}
In this paper, we motivate how the Hodge dual related with S-duality gives
the hidden symmetry in the moduli space of IIB string. Utilizing the static $%
\kappa $-symmetric Killing gauge, if we take the Hodge dual of the vierbeins
keeping the connection invariant, the duality of Maure-Cartan equations and
the equations of motion becomes manifest. Thus by twistly transforming the
vierbein, we can express the BPR currents as the Lax connections by a unique
spectral parameter. Then we construct the generators of the infinitesimal
dressing symmetry, the related symmetric algebra becomes the affine $%
gl(2,2|4)^{(1)}$, which can be used to find the classical $r$ matrix.
\end{abstract}

\section{Introduction}

Maldacena\cite{mald} proposed the AdS/CFT correspondence between the
classical theory of SUGRA on $AdS_{5}\otimes S^{5}$ in the bulk and the
quantum conformal SUSY Yang-Mills theory on the boundary.

In the last years, the existence of integrability structures on both sides
of the correspondence have been discussed in papers\cite%
{GSW,Alday,c1,c2,t1,t2,t3,t4,t5,t6,w1,w2,w3}. Bena, Polchinski and Roiban%
\cite{bpr} found an infinite set currents of classically conserved current
for the Green-Schwarz\cite{GS} string on $AdS_{5}\times S^{5},$ such that it
may be exact solvable. Dolan, Nappi and Witten\cite{DNW}\ have describe the
equivalences between this integrable structure and the Yangian symmetry of
the nonlocal currents as Bernards' paper\cite{bernard}.

For this infinite dimensional symmetry, the quantization procedure
will preserve it. But, to quantize such system, one has to handle
infinite number of the second class constraints given by the Dirac
quantization method. The ghost will appear, BRST method is needed
and the symmetry e.g. $\kappa $ symmetry will not be
manifest\cite{berkovits0001035}.

We try to disclose the complete symmetries of this classical system and to
construct the Lax-connection correspondently. Then the fundamental Poisson
bracket\cite{bracket} will give out the classical $r$ matrix\cite{crmatrix}.
Further, the quantization of $r$ matrix gives the quantum double $R$ matrix
which determines the scattering matrix. The amplitudes will be given in
terms of Bethe Ansatz method.

In this paper, we investigate the duality between the
Maurer-Cartan equations (MCE) and the equations of motion (EOM)
and obtain the Lax-matrix by using the twisted dual transformation
which represents a dreesing symmetry for GS string embedding into
$AdS_{5}\otimes S^{5}$. Physically, the invariance of
re-parametrization admits such twist transformation. However,
since on the MT action has imposed the conformal gauge, the action
is not invariant under re-parametrization. If we take the improved
stress-energy tensor the conformal symmetry will recovered as an
conformal affine symmetry. The full symmetric algebra will be
enlarged from $PSU\left( 2,2|4\right) $ into $gl(2,2|4)^{(1)}$.

\section{The Metsaev-Tseytlin action of Green-Shwarz superstring in $%
AdS^{5}\otimes S^{5}$ \cite{MT}}

The $AdS_{5}\otimes S^{5}$ is a coset space $\frac{SO(4,2)}{SO(4,1)}\otimes
\frac{SO(6)}{SO(5)}$. It also preserves the full supersymmetry of the SUGRA
and corresponds to the maximally supersymmetric background vacuum of IIB
SUGRA. Combining the bosonic $SO(4,2)\otimes SO(6)$ isometry symmetry with
the full supersymmetry, the symmetry turns to be the $PSU(2,2|4)$ acting on
the super coset space $\frac{PSU(2,2|4)}{SO(4,1)\otimes SO(5)}$. In what
follows, we adapt the conventions introduced by\cite{MT}:
\begin{equation*}
\begin{array}{rl}
a,b,c=0,1,\cdots ,4 &
\mbox{$so(4, 1)$ vector indices
($AdS_{5}$ tangent space)} \\
a^{\prime },b^{\prime },c^{\prime }=5,\cdots ,9 &
\mbox{$so(5)$
vector indices ($S^{5}$ tangent space)} \\
\hat{a},\hat{b},\hat{c}=0,1,\cdots ,9 &
\mbox{combination of
$(a, a^{\prime}), (b, b^{\prime}), (c, c^{\prime})$ ($D = 10$
vector indices)} \\
\alpha ,\beta ,\gamma ,\delta =1,\cdots ,4 &
\mbox{$so(4, 1)$
spinor indices ($AdS_{5}$)} \\
\alpha ^{\prime },\beta ^{\prime },\gamma ^{\prime },\delta ^{\prime
}=1,\cdots ,4 & \mbox{$so(5)$ spinor indices ($S^{5}$)} \\
\hat{\alpha},\hat{\beta},\hat{\gamma}=1,\cdots ,32 &
\mbox{$D
= 10$ Majorana-Weyl spinor indices} \\
I,J,K,L=1,2 & \mbox{SO(2) labels of the ${\cal N}=2$ two
sets of spinors}%
\end{array}%
\end{equation*}%
The generators of the $so(4,1)$ and $so(5)$ Clifford algebras are $4\times 4$
matrices $\gamma ^{a}$ and $\gamma ^{a^{\prime }}$
\begin{equation*}
\gamma ^{(a}\gamma ^{b)}=\eta ^{ab}=(-++++),\quad \gamma ^{(a^{\prime
}}\gamma ^{b^{\prime })}=\eta ^{a^{\prime }b^{\prime }}=(+++++).
\end{equation*}%
\begin{equation}
\hat{\gamma}^{a}\equiv \gamma ^{a},\quad \hat{\gamma}^{a^{\prime }}\equiv
i\gamma ^{a^{\prime }}
\end{equation}%
satisfying $(\gamma ^{a})^{\dagger }=\gamma ^{0}\gamma ^{a}\gamma ^{0}$, $%
(\gamma ^{a^{\prime }})^{\dagger }=\gamma ^{a^{\prime }}$ and the Majorana
condition is diagonal with respect to the two supercharges
\begin{equation}
\bar{Q}_{\alpha \alpha ^{\prime }I}\equiv (Q_{I}^{\beta \beta ^{\prime
}})^{\dagger }(\gamma ^{0})_{\alpha }^{\beta }\delta _{\alpha ^{\prime
}}^{\beta ^{\prime }}=-Q_{I}^{\beta \beta ^{\prime }}C_{\beta \alpha
}C_{\beta ^{\prime }\alpha ^{\prime }}.
\end{equation}%
Here $C=(C_{\alpha \beta })$ and $C^{\prime }=(C_{\alpha ^{\prime }\beta
^{\prime }})$ are the charge conjugation matrices of the $so(4,1)$ and $so(5)
$ Clifford algebras $Q_{\alpha \alpha ^{\prime }I}\equiv Q_{I}^{\beta \beta
^{\prime }}C_{\beta \alpha }C_{\beta ^{\prime }\alpha ^{\prime }}$. The
bosonic generators are antihermitean: $P_{a}^{\dagger }=-P_{a}$, $%
P_{a^{\prime }}^{\dagger }=-P_{a^{\prime }}$, $J_{ab}^{\dagger }=-J_{ab}$, $%
J_{a^{\prime }b^{\prime }}^{\dagger }=-J_{a^{\prime }b^{\prime }}$. The
SO(2) $2\times 2$ matrices are $\epsilon ^{IJ}=-\epsilon ^{JI}$, $\epsilon
^{12}=1$, and $s^{IJ}\equiv \mbox{diag}(1,-1)$.

The 10-dimensional $32 \times 32$ Dirac matrices $\Gamma^{\hat{a}}$ of $%
SO(9, 1)$ $(\Gamma^{( \hat{a}} \Gamma^{\hat{b} )} = \eta^{\hat{a} \hat{b}})$
and the corresponding charge conjugation matrix $\mathcal{C}$ can be
represented as
\begin{equation}
\Gamma^{a} = \gamma^{a} \otimes I \otimes \sigma_{1}, \quad
\Gamma^{a^{\prime}} = I \otimes \gamma^{a^{\prime}} \otimes \sigma_{2},
\quad \mathcal{C} = C \otimes C^{\prime} \otimes i \sigma_{2},
\end{equation}
where $I$ is the $4 \times 4$ unit matrix and $\sigma_{i}$ are the Pauli
matrices. The chirality is the eigenvalue of $\sigma_3$ in the last factor.

The generators of superalgebra $g$ $su(2,2|4)$ are divided into: a. the even
generators $\mathcal{B}$ which includes two pairs of translations and
rotations -- $(P_{a},J_{ab})\equiv (\mathfrak{h},\mathfrak{k})$ for $AdS_{5}$
and $(P_{a^{\prime }},J_{a^{\prime },b^{\prime }})\equiv (\mathfrak{h}%
^{\prime },\mathfrak{k}^{\prime })$ for $S^{5}$ respectively; b. the odd
generators $\mathcal{F}$ are the two $D=10$ Majorano-Weyl spinors $%
Q_{I}^{\alpha \alpha ^{\prime }}\equiv \mathcal{F}_{I}$.

The commutation relations for the generators $T_{A}=(P_{a},P_{a^{\prime
}};J_{ab},J_{a^{\prime }b^{\prime }}|Q_{\alpha \alpha ^{\prime }I})$ $\equiv
(\mathfrak{h},\mathfrak{h}^{\prime };\mathfrak{k},\mathfrak{k}^{\prime }|%
\mathcal{F}_{I})$ $\equiv (\hat{\mathfrak{h}};\hat{\mathfrak{k}}|\mathcal{F}%
) $ are
\begin{equation}
\lbrack J_{ab},J_{cd}]=\eta _{bc}J_{ad}+3\mbox{ terms}\ ,\quad \lbrack
\mathfrak{h},\mathfrak{h}]\subset \mathfrak{h}\ ,
\end{equation}%
\begin{equation}
\lbrack J_{a^{\prime }b^{\prime }},J_{c^{\prime }d^{\prime }}]=\eta
_{b^{\prime }c^{\prime }}J_{a^{\prime }d^{\prime }}+3\mbox{
terms},\quad \lbrack \mathfrak{h}^{\prime }\ ,\mathfrak{h}^{\prime }]\subset
\mathfrak{h}^{\prime }\ ,
\end{equation}%
\begin{equation}
\lbrack P_{a},P_{b}]=J_{ab}\ ,\quad \lbrack \mathfrak{k},\mathfrak{k}%
]\subset \mathfrak{h}
\end{equation}%
\begin{equation}
\lbrack P_{a^{\prime }},P_{b^{\prime }}]=-J_{a^{\prime }b^{\prime }}\ ,\quad
\lbrack \mathfrak{k}^{\prime },\mathfrak{k}^{\prime }]\subset \mathfrak{h}%
^{\prime }
\end{equation}%
\begin{equation}
\lbrack P_{a},J_{bc}]=\eta _{ab}P_{c}-\eta _{ac}P_{b}\ ,\quad \lbrack
\mathfrak{h},\mathfrak{k}]\subset \mathfrak{k}\ ,
\end{equation}%
\begin{equation}
\lbrack P_{a^{\prime }},J_{b^{\prime }c^{\prime }}]=\eta _{a^{\prime
}b^{\prime }}P_{c^{\prime }}-\eta _{a^{\prime }c^{\prime }}P_{b^{\prime
}},\quad \lbrack \mathfrak{h}^{\prime },\mathfrak{k}^{\prime }]\subset
\mathfrak{k}^{\prime }.
\end{equation}%
\begin{equation}
\lbrack Q_{I},P_{a}]=-\frac{i}{2}\epsilon _{IJ}Q_{J}\gamma _{a},\quad
\lbrack \mathcal{F}^{I},\mathfrak{k}]\subset \epsilon _{IJ}\mathcal{F}^{J},
\end{equation}%
\begin{equation}
\lbrack Q_{I},P_{a^{\prime }}]=\frac{1}{2}\epsilon _{IJ}Q_{J}\gamma
_{a^{\prime }},\quad \lbrack \mathcal{F}^{I},\mathfrak{k}^{\prime }]\subset
\epsilon _{IJ}\mathcal{F}^{J},
\end{equation}%
\begin{equation}
\lbrack Q_{I},J_{ab}]=-\frac{1}{2}Q_{I}\gamma _{ab},\quad \lbrack \mathcal{F}%
^{I},\mathfrak{h}]\subset \mathcal{F}^{I},
\end{equation}%
\begin{equation}
\lbrack Q_{I},J_{a^{\prime }b^{\prime }}]=-\frac{1}{2}Q_{I}\gamma
_{a^{\prime }b^{\prime }},\quad \lbrack \mathcal{F}^{I},\mathfrak{h}^{\prime
}]\subset \mathcal{F}^{I}.
\end{equation}%
\begin{eqnarray}
\{Q_{\alpha \alpha ^{\prime }I},Q_{\beta \beta ^{\prime }J}\} &=&\delta
_{IJ} \left[ -2iC_{alpha^{\prime }\beta ^{\prime }}(C\gamma ^{a})_{\alpha
\beta }P_{a}+2C_{\alpha \beta }(C^{\prime }\gamma ^{a^{\prime }})_{\alpha
^{\prime }\beta ^{\prime }}P_{a^{\prime }}\right]  \notag \\
&+&\epsilon _{IJ}\left[ C_{\alpha ^{\prime }\beta ^{\prime }}(C\gamma
^{ab})_{\alpha \beta }J_{ab}-C_{\alpha \beta }(C^{\prime }\gamma ^{a^{\prime
}b^{\prime }})_{\alpha ^{\prime }\beta ^{\prime }}J_{a^{\prime }b^{\prime }}%
\right] \ , \\
&&\qquad \qquad \qquad \qquad \lbrack \mathcal{F},\mathcal{F}]\subset
\mathcal{B}\ .  \notag
\end{eqnarray}

The left-invariant Cartan 1-forms
\begin{equation}
L^{A}=dX^{M}L_{M}^{A},\quad X^{M}=(x,\theta )
\end{equation}%
are given by
\begin{equation}
G^{-1}dG=L^{A}T_{A}\equiv L^{a}P_{a}+L^{a^{\prime }}P_{a^{\prime }}+\frac{1}{%
2}L^{ab}J_{ab}+\frac{1}{2}L^{a^{\prime }b^{\prime }}J_{a^{\prime }b^{\prime
}}+L^{\alpha \alpha ^{\prime }I}Q_{\alpha \alpha ^{\prime }I}\ ,
\label{cartanform}
\end{equation}%
where $G=G(x,\theta )$ is a coset representative in $PSU(2,2|4)$, e.g. by
using the S-gauge given by\cite{Met} or the KRP gauge\cite%
{kallosh9805217,pesando9808020} used by Roiban and Siegel\cite{RS}.

We call $\hat{H}(H,H^{\prime })$, $\hat{K}(K,K^{\prime })$, $K_{F}$ forms
respectively for the Cartan connections $L^{ab}$ and $L^{a^{\prime
}b^{\prime }}$, the super-beins $L^{\hat{a}\hat{\alpha}}$ including
funf-beins $L^{a}$ and $L^{a^{\prime }}$ and the 2-spinor 16-beins $%
L^{\alpha \alpha ^{\prime }I}$. They satisfy the Maurer-Cartan (MC)
equation, i.e. the structure equation of basic one forms on the superspace $%
\frac{SU(2,2|4)}{SO(4,1)\otimes SO(5)}$
\begin{equation}
d(G^{-1}dG)+(G^{-1}dG)\wedge (G^{-1}dG)=0.
\end{equation}%
Then the super Gauss equations of the induced curvatures $F^{ab}$ and $%
F^{a^{\prime }b^{\prime }}$ defined by $F=dH+H\wedge H$ are
\begin{eqnarray}
F^{ab} &\equiv &dL^{ab}+L^{ac}\wedge L^{cb}=-L^{a}\wedge L^{b}+\epsilon ^{IJ}%
\bar{L}^{I}\gamma ^{ab}\wedge L^{J}, \\
F^{a^{\prime }b^{\prime }} &\equiv &dL^{a^{\prime }b^{\prime }}+L^{a^{\prime
}c^{\prime }}\wedge L^{c^{\prime }b^{\prime }}=L^{a^{\prime }}\wedge
L^{b^{\prime }}-\epsilon ^{IJ}\bar{L}^{I}\gamma ^{a^{\prime }b^{\prime
}}\wedge L^{J}.
\end{eqnarray}%
The super Coddazi equation for the funf-beins are
\begin{equation}
dL^{a}+L^{b}\wedge L^{ba}=-iL^{I}\gamma ^{a}\wedge L^{I},\quad dL^{a^{\prime
}}+L^{b^{\prime }}\wedge L^{b^{\prime }a^{\prime }}=L^{I}\gamma ^{a^{\prime
}}\wedge L^{I},  \label{Coddazi5}
\end{equation}%
and the super Coddazi equation for the spinor 16-beins are
\begin{equation}
dL^{I}-\frac{1}{4}\gamma ^{ab}L^{I}\wedge L^{ab}-\frac{1}{4}\gamma
^{a^{\prime }b^{\prime }}L^{I}\wedge L^{a^{\prime }b^{\prime }}=-\frac{1}{2}%
\gamma ^{a}\epsilon ^{IJ}L^{J}\wedge L^{a}+\frac{1}{2}\epsilon ^{IJ}\gamma
^{a^{\prime }}L^{J}\wedge L^{a^{\prime }}\ .  \label{gc3}
\end{equation}

In the super Gauss equations and the Coddazi equations, the terms on the
left hand side are the usual gauge covariant exterior derivative $d+H\wedge $%
, while the right hand side include the contributions of curvature and
torsion by the fermions.

To embed the IIB superstring into the super coset space
$\mathcal{M}$, we should pull back the Cartan form down to the
world sheet $\Sigma (\sigma ,\tau )$ as
\begin{equation}
L^{A}=L_{M}^{A}dx^{M}=L_{M}^{A}\partial _{i}x^{M}d\sigma
^{i}=L_{i}^{A}d\sigma ^{i}\equiv L_{1}^{A}d\tau +L_{2}^{A}d\sigma \ .
\end{equation}%
Then the MC 1-form becomes
\begin{equation}
G^{-1}\partial _{i}G=L_{i}^{A}P_{A}=L_{i}^{a}P_{a}+L_{i}^{a^{\prime
}}P_{a^{\prime }}+\frac{1}{2}(L_{i}^{ab}J_{ab}+L_{i}^{a^{\prime }b^{\prime
}}J_{a^{\prime }b^{\prime }})+L_{i}^{\alpha \alpha ^{\prime }I}Q_{\alpha
\alpha ^{\prime }I}\ ,
\end{equation}%
and e.g. the super Coddazi equations for the vector 5-beins (\ref{Coddazi5})
become
\begin{eqnarray}
\epsilon ^{ij}(\partial _{i}L_{j}^{a}+L_{i}^{ab}L_{j}^{b})+i\epsilon ^{ij}%
\bar{L}_{i}^{I}\gamma ^{a}L_{j}^{I} &=&0\ ,  \label{gc1} \\
\epsilon ^{ij}(\partial _{i}L_{j}^{a^{\prime }}+L_{i}^{a^{\prime }b^{\prime
}}L_{j}^{b^{\prime }})-\epsilon ^{ij}\bar{L}_{i}^{I}\gamma ^{a^{\prime
}}L_{j}^{I} &=&0\ .  \label{gc2}
\end{eqnarray}%
The MC equations for the vierbeins describes the \textbf{geometric} behavior
for the embedding of the type IIB string worldsheet into the target space $%
AdS_{5}\times S^{5}$.

Now turn to the string \textbf{dynamics}. the $AdS_{5}\otimes S^{5}$ GS
superstring action is given as $\frac{SU(2,2|4)}{SO(4,1)\otimes SO(5)}$
superspace sigma model\cite{MT,Met}.
\begin{equation}
I=-\frac{1}{2}\int_{\partial M_{3}}d^{2}\sigma \sqrt{g}%
g^{ij}(L_{i}^{a}L_{j}^{a}+L_{i}^{a^{\prime }}L_{j}^{a^{\prime
}})+i\int_{M_{3}}s^{IJ}(L^{a}\wedge \bar{L}^{I}\gamma ^{a}\wedge
L^{J}+iL^{a^{\prime }}\wedge \bar{L}^{I}\gamma ^{a^{\prime }}\wedge L^{J}).
\label{action}
\end{equation}%
This action is invariant with respect to the local $\kappa $-transformations
in terms of $\delta x^{a}\equiv \delta X^{M}L_{M}^{a}$, $\delta x^{a^{\prime
}}\equiv \delta X^{M}L_{M}^{a^{\prime }}$, $\delta \theta ^{I}\equiv \delta
X^{M}L_{M}^{I}$
\begin{eqnarray}
&&\delta _{\kappa }x^{a}=0,\quad \delta _{\kappa }x^{a^{\prime }}=0,\quad
\delta _{\kappa }\theta ^{I}=2(L_{i}^{a}\gamma ^{a}-iL_{i}^{a^{\prime
}}\gamma ^{a^{\prime }})\kappa ^{iI} \\
&&\delta _{\kappa }(\sqrt{g}g^{ij})=-16i\sqrt{g}(P_{-}^{jk}\bar{L}%
_{k}^{1}\kappa ^{i1}+P_{+}^{jk}\bar{L}_{k}^{2}\kappa ^{i2})\ .
\end{eqnarray}%
Here $P_{\pm }^{ij}\equiv \frac{1}{2}(g^{ij}\pm \frac{1}{\sqrt{g}}\epsilon
^{ij})$ are the projection operators, and $16$-component spinor $\kappa ^{iI}
$ (the corresponding $32$-component spinor has opposite chirality to that of
$\theta $) satisfy the (anti) self duality constraints
\begin{equation}
P_{-}^{ij}\kappa _{j}^{1}=\kappa ^{i1},\quad P_{+}^{ij}\kappa
_{j}^{2}=\kappa ^{i2},
\end{equation}%
which can be written as $\frac{1}{\sqrt{g}}\epsilon ^{ij}\kappa
_{j}^{1}=-\kappa ^{i1}$, $\frac{1}{\sqrt{g}}\epsilon ^{ij}\kappa
_{j}^{2}=\kappa ^{i2}$, i.e. $\frac{1}{\sqrt{g}}\epsilon ^{ij}\kappa
_{j}^{I}=-S^{IJ}\kappa ^{iJ}$.

From the variation of action (\ref{action}), the equations of motion (EOM)
are obtained\cite{MT}
\begin{eqnarray}
\sqrt{g}g^{ij}(\bigtriangledown _{i}L_{j}^{a}+L_{i}^{ab}L_{j}^{b})+i\epsilon
^{ij}s^{IJ}\bar{L}_{i}^{I}\gamma ^{a}L_{j}^{J} &=&0\ ,  \label{EM1} \\
\sqrt{g}g^{ij}(\bigtriangledown _{i}L_{j}^{a^{\prime }}+L_{i}^{a^{\prime
}b^{\prime }}L_{j}^{b^{\prime }})-\epsilon ^{ij}s^{IJ}\bar{L}_{i}^{I}\gamma
^{a^{\prime }}L_{j}^{J} &=&0\ ,  \label{EM2} \\
(\gamma ^{a}L_{i}^{a}+i\gamma ^{a^{\prime }}L_{i}^{a^{\prime }})(\sqrt{g}%
g^{ij}\delta ^{IJ}-\epsilon ^{ij}s^{IJ})L_{j}^{J} &=&0\ ,  \label{EM3}
\end{eqnarray}%
where $\bigtriangledown _{i}$ is the $g_{ij}$-covariant derivative on the
worldsheet $\Sigma (\sigma ,\tau )$.

\section{The Hodge dual and duality between MCE and EOM}

The equations of motion (\ref{EM1}) and (\ref{EM2}) can be rewritten as
\begin{eqnarray}
g^{ij}(\partial _{i}(\sqrt{g}L_{j}^{a})+L_{i}^{ab}L_{j}^{b})+i\epsilon
^{ij}S^{IJ}\bar{L}_{i}^{I}\gamma ^{a}L_{j}^{J} &=&0\ ,  \label{eq1} \\
g^{ij}(\partial _{i}(\sqrt{g}L_{j}^{a^{\prime }})+L_{i}^{a^{\prime
}b^{\prime }}L_{j}^{b^{\prime }})-\epsilon ^{ij}S^{IJ}\bar{L}_{i}^{I}\gamma
^{a^{\prime }}L_{i}^{J} &=&0\ ,  \label{eq2}
\end{eqnarray}%
where we have used
\begin{equation}
\bigtriangledown _{i}L^{\hat{a}i}=\frac{1}{\sqrt{g}}\partial _{i}(\sqrt{g}L^{%
\hat{a}i})\ .
\end{equation}

In order to disclose the duality between the MCE and the EOM, we first
describes the Hodge dual of bosonic and fermionic forms.

As usual, the Hodge dual of the coordinates of world-sheet is given by
\begin{eqnarray}
^{\ast }(dz^{i}) &=&\frac{-1}{\sqrt{g}}\epsilon ^{ij}dz_{j},\quad
(dz^{1})=d\tau ,\quad (dz^{2})=d\sigma \ ,  \notag \\
\epsilon _{12} &=&-\epsilon _{21}=\epsilon ^{21}=-\epsilon ^{12}=1\ .
\end{eqnarray}%
So the even beins simply become%
\begin{equation}
L^{i\hat{a}}\longleftrightarrow \,^{\ast }L^{i\hat{a}}\equiv \left( \,^{\ast
}L^{\hat{a}}\right) ^{i}=-\frac{\epsilon ^{ij}}{\sqrt{g}}L_{j}^{\hat{a}}\
,\quad L_{i}^{\hat{a}}\longleftrightarrow \,^{\ast }L_{i}^{\hat{a}}\equiv {%
\epsilon _{ij}\sqrt{g}}L^{j\hat{a}}\ .
\end{equation}%
As for the odd part, i.e. the spinors in the static $\kappa $ symmetric
Killing gauge\cite{kallosh9805217,pesando9808020,killing}, we have
\begin{eqnarray}
P_{-}^{ij}L_{j}^{1} &=&L^{1i}\ ,\quad P_{+}^{ij}L_{j}^{2}=L^{2i}\ ,  \notag
\\
P_{-}\kappa ^{1} &=&\kappa ^{1}\ ,\quad P_{+}\kappa ^{2}=\kappa ^{2}\ ,\quad
\,^{\ast }\theta ^{I}\equiv S^{IJ}\theta ^{J}\ ,  \label{killing}
\end{eqnarray}%
here $\kappa \left( x\right) ,\theta \left( x\right) $\ are $\pm $\
eigenvector of local $\Gamma ^{11}\left( x\right) $ which satisfy Killing
equation\cite{lu9805151}. This implies that all the local tangent vectors
and eigen spinors are covariant \textbf{in the same local gauge}, which is
covariantly moving on $AdS_{5}$. And this covariance is enabled by the pure
geometrical behavior of $AdS^{5}\otimes S^{5}$ in SUGRA\cite{kallosh9805041}%
. Then the MCE (\ref{gc1}) and (\ref{gc2}) can be rewritten as
\begin{eqnarray}
\partial _{i}(\sqrt{g}\;^{\ast }L^{ia})+L_{i}^{ab}\sqrt{g}\;^{\ast
}L^{ib}-i\epsilon ^{ij}\bar{L}_{i}^{I}\gamma ^{a}L_{j}^{I} &=&0\ ,
\label{MC1} \\
\partial _{i}(\sqrt{g}\;^{\ast }L^{ia^{\prime }})+L_{i}^{a^{\prime
}b^{\prime }}\sqrt{g}\;^{\ast }L^{ib^{\prime }}+\epsilon ^{ij}\bar{L}%
_{i}^{I}\gamma ^{a^{\prime }}L_{j}^{I} &=&0\ .  \label{MC2}
\end{eqnarray}

Applying the above transformations, it is easy to find that the EOM (\ref%
{EM1}) and (\ref{EM2}) are the dual of the 5-bein super Codazzi equations (%
\ref{MC1}) and (\ref{MC2}) respectively.

It is clear that the GS string action is invariant under the above dual
transformation. There exists no dual between MC eq.({\ref{gc3}) and EOM eq.(%
\ref{EM3}), because the $L^{I}$ only appears in the Wess-Zumino-Witten term
and has no dynamical contribution to the action. Under the dual
transformation, the 3rd EOM }$\left( \ref{EM3}\right) ${\ changes into
\begin{equation}
(\gamma ^{a}\;^{\ast }L_{i}^{a}+i\gamma ^{a^{\prime }}\;^{\ast
}L_{i}^{a^{\prime }})(\sqrt{g}g^{ij}\delta ^{IJ}-\epsilon
^{ij}s^{IJ})L_{j}^{J}=0\ .
\end{equation}%
Namely, only the first factor takes the dual form. }

{For the $L^{\hat{a}\hat{b}}$, it does not change under duality since it is
not dynamical and does not appear in the GS string action. }

Remark: this duality is the generalization of the usual bosonic Hodge dual
between the dynamical first fundamental form (metric 1-form) of the
pseudosphere and its geometric 2nd fundamental form. For the pseudo-sphere
with negative constant curvature
\begin{equation}
d\omega _{12}=-k\omega _{1}\wedge \omega _{2}\ ,  \label{omga2}
\end{equation}%
where $\omega _{12}$ is connection 1-form and the constant curvature $k=-1/%
\sqrt{g}$. The metric 1-form is $ds^{2}=d\omega _{1}^{2}+d\omega _{2}^{2}$.
Under the \textbf{same moving frame} i.e. the \textbf{same gauge}, one has
\begin{equation}
\omega ^{i}=\epsilon ^{ij}\;^{\ast }d\omega _{3j}\ .  \label{omga1}
\end{equation}%
With the help of eq.(\ref{omga1}), the eq.(\ref{omga2}) changes into
\begin{equation}
d\omega _{12}=-\omega _{13}\wedge \omega _{23}\ ,
\end{equation}%
which is the 3rd component of the Maure-Cardan equation. Since the
connection form $\omega _{12}$ which is shared by both sides of
duality, will be changed in the same way under same
gauge\cite{hou,hhbook}. So while we take the duality for other two
components (Coddazi eq.) to interchange the MCE into the EOM, the
$\omega _{12}$ will be the same. The geodesic motion on
pseudo-sphere gives dynamical Sine-Gordon equation of the angle
between asymptetic lines. The image of the normal line of
pseudo-sphere gives the non-linear $\sigma $ model on the sphere.
For the non-linear $\sigma $ model
on $AdS_{5}$, the conformal metric is the Poincar\'{e} metric $\frac{%
\sum_{i=1}^{3}dx_{i}^{2}+dr^{2}}{r^{2}}$ as the higher dimensional
generalization for the metric of pseudo-sphere. The dual of (\ref{EM1}) and (%
\ref{EM2}) to eq.(\ref{MC1}) and (\ref{MC2}) can be considered as the
generalization of such duality in $AdS_{5}\otimes S^{5}$ with the $\kappa $
symmetric gauge.

\section{The twisted dual and integrability}

Now we introduce the twisted dual transformation of vierbein as follows. The
duality discussed in previous section will be included as a special case of
it. On the world sheet $\Sigma (\sigma ,\tau )$, it is the
re-parametrization transformations along the two directions of the positive
and negative light-cone $\tau \pm \sigma $ with the scale factors $\lambda
=e^{2\phi }$ and $\lambda ^{-1}$ correspondently, the even vierbein forms $%
L^{\hat{a}}$ will be Lorentz rotate by $\pm 2\phi $ oppositely,
and the odd vierbein forms $L^{I}$ will rotate oppsitely by $\pm
\varphi $ together with $\theta ^{J}$\ and $\kappa ^{I}$. All
vierbeins are rotate around same axis, the normal line of
$AdS_{5}$ surface \textbf{in the same gauge} especially in Killing
gauge.

For the even funf-bein form part, we have
\begin{eqnarray}
\mathcal{L}^{i\hat{a}}\left( \lambda \right) &\equiv &\exp ^{2\varphi
}P_{+}^{ij}L_{j}^{\hat{a}}+\exp ^{-2\varphi }P_{-}^{ij}L_{j}^{\hat{a}}
\notag \\
&=&\frac{1}{2}(\lambda +\lambda ^{-1})L^{i\hat{a}}+\frac{1}{2}(\lambda
-\lambda ^{-1})\;^{\ast }L^{i\hat{a}},
\end{eqnarray}%
here $\lambda =\exp ^{2\varphi }$.

The transformations of odd vierbein are%
\begin{eqnarray}
\mathcal{L}^{i1}\left( \lambda \right)  &\equiv &\lambda ^{-\frac{1}{2}%
}L_{j}^{1}=e^{\varphi }P_{-}^{ij}L_{j}^{1},  \label{1} \\
\mathcal{L}^{i2}\left( \lambda \right)  &\equiv &\lambda ^{\frac{1}{2}%
}L_{j}^{2}=e^{-\varphi }P_{+}^{ij}L_{j}^{2},  \label{2}
\end{eqnarray}%
here the Killing gauge $\left( \ref{killing}\right) $ are used. Combined $%
\left( \ref{1}\right) $ and $\left( \ref{2}\right) ,$ it yields
\begin{equation}
\mathcal{L}^{iI}\left( \lambda \right) =\exp ^{\varphi
}P_{+}^{ij}L_{j}^{I}+\exp ^{-\varphi }P_{-}^{ij}L_{j}^{I},  \label{fer1}
\end{equation}%
\begin{equation}
\bar{\mathcal{L}}^{iI}\left( \lambda \right) =\exp ^{\varphi }P_{+}^{ij}\bar{%
L}_{j}^{I}+\exp ^{-\varphi }P_{-}^{ij}\bar{L}_{j}^{I}.  \label{fer2}
\end{equation}%
Notice that the re-parametrization invariance of action implies the loop
group symmetry, while the WZW term as a 2-cocycle, further gives the central
extension. Here the $GL(1)\otimes GL(1)$ supplies an axial symmetry $A$ of
the extended $\mathcal{N}=2$ SUSY, missing in the PSU$\left( 2,2|4\right) $
GS string action in Ref.\cite{MT}, and the $U(1)$ S-duality of $\frac{SL(2,R)%
}{U(1)}\sim \frac{SU(1,1)}{U(1)}$ in super-gravity. The dual twist breaks
the invariance of the action in conformal gauge, $T_{a}^{a}\neq 0$
superficially. But, one can recover it by using the improved stress-energy
tensor with contributions from the $U(1)\times U(1)$ fields which correspond
to the central and grade derivative operators in affine algebra\cite{b244}.
This is not the symmetry of MT action, actually it is the hidden symmetry in
the moduli space, which is described by the continuous spectral parameter $%
\lambda .$

Now we can construct the Lax connection $A_{i}\left( \lambda \right) $ with
the spectral parameter $\lambda $ as
\begin{eqnarray}
A_{i}(\lambda ) &=&H+\mathcal{K}\left( \lambda \right) +\mathcal{F}\left(
\lambda \right)  \notag \\
&=&\frac{1}{2}L_{i}^{\hat{a}\hat{b}}J_{\hat{a}\hat{b}}+\mathcal{L}_{i}^{\hat{%
a}}\left( \lambda \right) P_{\hat{a}}+\mathcal{L}_{i}^{\alpha \alpha
^{\prime }I}\left( \lambda \right) Q_{\alpha \alpha ^{\prime }I}  \notag \\
&=&\frac{1}{2}(L_{i}^{ab}J_{ab}+L_{i}^{a^{\prime }b^{\prime }}J_{a^{\prime
}b^{\prime }})+\frac{1}{2}(\lambda +\lambda
^{-1})(L_{i}^{a}P_{a}+L_{i}^{a^{\prime }}P_{a^{\prime }})  \notag \\
&&+\frac{1}{2}(\lambda -\lambda ^{-1})\Big[\;^{\ast
}(L^{a})_{i}P_{a}+\;^{\ast }(L^{a^{\prime }})_{i}P_{a^{\prime }}\Big]  \notag
\\
&&+\lambda ^{-\frac{1}{2}}L_{i}^{\alpha \alpha ^{\prime }1}Q_{\alpha \alpha
^{\prime }1}+\lambda ^{\frac{1}{2}}L_{i}^{\alpha \alpha ^{\prime
}2}Q_{\alpha \alpha ^{\prime }2}\ ,
\end{eqnarray}%
which looks like the original Cartan form with beins replaced by $\mathcal{L}%
\left( \lambda \right) .$ Such an O(2) transformation, should be defined in
the same covariantly shifted moving frame, (the same gauge) with \textbf{%
covariant constant} $N(x)$. Thus the $H$ including in the covariant
derivative will not be twisted. Obviously if $\lambda =1,\phi =0,$ it is the
original Cartan form $\left( \ref{cartanform}\right) .$ On the "wick
rotated" worldsheet we may take
\begin{equation}
\lambda =\exp ^{2\varphi }=i.
\end{equation}%
Then
\begin{equation}
\mathcal{L}^{i\hat{a}}=i\epsilon ^{ij}L_{j}^{\hat{a}},  \label{dt1}
\end{equation}%
Similarly%
\begin{equation}
\mathcal{L}^{i1}=(-i)^{\frac{1}{2}}L^{i1},\quad \mathcal{L}^{i2}=i^{\frac{1}{%
2}}L^{i2},  \label{dt2}
\end{equation}%
\begin{equation}
\bar{\mathcal{L}}^{i1}=(-i)^{\frac{1}{2}}\bar{L}^{i1},\quad \bar{\mathcal{L}}%
^{i2}=i^{\frac{1}{2}}\bar{L}^{i2},  \label{dt3}
\end{equation}%
\begin{equation}
i.e.\mathcal{L}^{\hat{a}}=\,^{\ast }L^{\hat{a}}\text{ and }%
L^{I}=S^{IJ}\,^{\ast }L^{J},
\end{equation}%
here i appears, from the difference of sign of Hodge star in
$\mathbb{M}_{2}$ and in $\mathbb{E}_{2}.$ Thus the vierbein
$\mathcal{L}(i)$ becomes simply the Hodge dual of original
vierbein on Euclidean world sheet. And it implies the dual
symmetry of the MCE and the EOM. It is obvious that the Lax
connections $A_{i}(\lambda )$ satisfy the zero curvature (flat
connection) condition:
\begin{equation}
\partial _{i}A_{j}(\lambda )-\partial _{j}A_{i}(\lambda )+[A_{i}(\lambda
),A_{j}(\lambda )]=0,
\end{equation}%
as the linear combination of MCE and EOM i.e. the system is integrable, and
we may introduce the transfer matrices $U\left( \lambda ,\sigma \right) $
\begin{equation}
\partial _{i}U\left( \lambda ,\sigma \right) =A_{i}\left( \lambda ,\sigma
\right) U\left( \lambda ,\sigma \right) .  \label{u}
\end{equation}

\section{Infinite set of conserved currents}

The EOM (\ref{EM1},\ref{EM2}) is the conservation law of the Noether
currents with respect to the local transitive operator $\mathbb{T}^{a}$. In
the KPR gauge, let the normal line $N\left( x\right) $, $N^{2}\left(
x\right) =1$ (the normalized $Y\left( x\right) $ in Ref.\cite{RS}) be $e_{6}$%
. These generators $\mathbb{T}^{a}\sim T^{6a}\in SO(4,2)$ in the right
\textquotedblleft body axis\textquotedblright\ becomes the translation
generated by Killing vectors along the same direction, here $e_{a}\left(
a=1\cdots 5\right) $ are the tangent vectors on $AdS_{5}$.

The existence of transfer matrix $U\left( \lambda ,x\right) $ implies that
Noether currents generated by $U^{-1}\left( \lambda ,x\right) \mathbb{T}%
^{a}U\left( \lambda ,x\right) $ is conserved. We omit the detail of the
proof. It's essentially the generalization of Ref.\cite{hou,bouyale,houmath}%
. Here besides%
\begin{eqnarray}
\mathcal{D}_{i}\left( \lambda \right) U\left( \lambda \right)  &=&0, \\
\mathcal{D}_{i}\left( \lambda \right)  &=&\partial _{i}+A_{i}\left( \lambda
\right) ,  \label{58}
\end{eqnarray}%
the covariant constancy of
\begin{eqnarray}
N\left( \lambda ;x\right)  &=&U^{-1}\left( \lambda \right) N\left(
1;x\right) U\left( \lambda \right) , \\
\mathcal{D}_{i}\left( \lambda \right) N\left( \lambda ;x\right)  &=&0,
\label{59}
\end{eqnarray}%
here $\mathcal{D}_{i}\left( \lambda \right) =\left( \partial
_{i}+[A_{i}\left( \lambda \right) ,]\right) $ in adjoint
representation is used.

Notice: Three different covariant derivatives

\begin{enumerate}
\item covariant derivative $\nabla _{i}$ on the world sheet (eq.(\ref{EM1}-%
\ref{EM3}))

\item gauge covariant exterior derivative $d+H\wedge $ on $S^{5}$ and (or)
on $AdS_{5}$ surface.

\item covariant derivative D$_{i}$ (\ref{58}) (\ref{59}) in the flat
\textquotedblleft zero curvature\textquotedblright\ 6-dim space.
\end{enumerate}

\section{Outlook}

Soon later , after the Yale preprint\cite{bouyale} Ueno\cite{u1}
pointed out that $U^{-1}TU,T\in \mathfrak{g}$ is the infinitesimal
\textbf{dressing transformation}. It is the generator of dressing
transformation\cite{rh} which is related with finite
\textbf{Riemann Hilbret} problem $U^{-1}gU,g\in G.$ The Riemann
Hilbret problem with poles and zeros will determine the left and
right part of Leznov, Sovileevs soliton solution. So Riemann
Hilbret factorization gives the holomorphic and antiholomorphic of
left and right moving part of strings and branes. They combine
into amplitude as that in CFT how correlation function is obtained
from conformal block. Furthermore, the \textbf{Possion Lie
structure}\cite{crmatrix} of this affine dressing group is given.
The soliton generation operator will be constructed by
exponentiate the $r_{\pm }$ matrix of the classical double, it is just the $%
\pm $ frequency part of the vertex operator of principle realization of
affine Kac Moody algebra\cite{kac}.

We will find classical solitonic solution of IIB string and brane. The
\textbf{loop parameter} describes their moduli space. The loop parameter $%
\lambda $ also characterized the constant change of phase or
length of the complexified $O(2)$ rotation around the normal line
$N$, which is the well know covariant constant field in SUGRA.
This $O(2)$ is the S duality $U(1)$ in $\frac{SU(1,1)}{U(1)}.$
This can be generalize into the maximally symmetry $AdS_{7}\times
S^{4}$ , $AdS_{4}\times S^{7}$ of M theory, with pure geometrical
behavior. And then transform to other dimensions by T duality.

The Hodge twist transform is the case of unequal fermion parameter $\alpha
\neq \bar{\alpha}$ in Witten's topological nonlinear $\sigma \ $model$,$
thus $\lambda $ gives different moduli of IIA, IIB\cite{mirror}. At last $%
\lambda $ will describe the affinization of the R symmetry and dilation of
the dimension.

Studying the \textbf{Poisson structure}, Dolan find the so called Kac Moody
alg. of principle chiral model. She points out that it is the loop parameter
expansion of the Poisson structure of the transformation in Ref.\cite{hou}.
In fact, it's the fundamental Poisson bracket given by Faddeev%
\begin{equation}
\{U(\lambda ),\otimes U(\mu )\}=[r(\lambda -\mu ),U(\lambda )\otimes U(\mu
)],
\end{equation}%
and the Poisson Lie bracket%
\begin{equation}
\{L(\varphi ),\otimes L(\psi )\}=[r(\varphi -\psi ),L(\varphi )\otimes
\mathbf{1}+\mathbf{1}\otimes L(\psi )],  \label{plb}
\end{equation}%
here $L$ is the $A_{0}$ component in (\ref{u})$.$

In fact, the $r$-matrix may be derived from the symplectic form given by the
action with WZW term\cite{AA}. It is just the classical algebra. The quantum
version of (\ref{plb}) is given by Drinfeld, Jimbo, Faddeev, Reshetikin and
Takhtajan\cite{R1,R2,R3},
\begin{equation}
R_{12}(\phi -\psi )L_{1}(\phi )L_{2}(\psi )=L_{2}(\psi )L_{1}(\phi
)R_{12}(\phi -\psi ).
\end{equation}

In a further paper (hep-th/0406250) we will give the gauged form
of Roiban-Siegel action\cite{RS}, corresponding to the embedding
of IIB string in $AdS_{5}\times S^{5}.$ Its bosonic part is the
the gauged WZW action for affine Toda theory\cite{Babelon}. Thus,
the the dressing transformation gives all the soliton solution,
i.e. the moduli space of the string background states. After the
quantization, it will become quantum $R$ and $L$
matrices\cite{houyang}\ for the quantum Affine Toda. The
scattering amplitude can also be obtained as Ref.\cite{abz}. The
correlation function of vertex operators is given by the double
scaling limit of the quantum Virasoso and W algebra\cite{frenkel},
which is equivalent to the q-deformation of Yangian double with
centre\cite{khoroshkin} at the critical level\cite{houzhao}. Thus
the dressing symmetry revealed in this paper discloses that both
the classical and quantum integrable structure of IIB GS string on
$AdS_{5}\times S^{5}$ is given by the Affine Toda system. e.g. why
the Seiberg-Witten curve is given by the spectral determinant of
affine Toda.

A chinese version of this paper as part of the Doctor Dissertation of
Chuan-Hua Xiong will be published in ACTA PHYSICA SINICA in chinese
(received on March 9).

\section*{Acknowledgments}

We are grateful to Bo-Yuan Hou, Xing-Chang Song and Kang-Jie Shi for helpful
discussions.

Note added: Thanks to Xiang-Mao Ding for attract our attention to A. M.
Polyakov's paper\cite{polykov}, where the graded opposite rotation with the
spectral parameter $\lambda $ and zero curvature relation equation has been
given.


\begin{thebibliography}{99}
\bibitem{mald} J. M. Maldacena, \textquotedblleft The Large N Limit of
Superconformal Field Theories and Supergravity\textquotedblright , \textit{%
Adv. Theor. Math. Phys.} \textbf{2} (1998) 231-252; \textit{Int. J. Theor.
Phys.} \textbf{38} (1999) 1113-1133, [arXiv:hep-th/9711200].

\bibitem{GSW} Niklas Beisert, \textit{JHEP}\textbf{\ 0309} (2003) 062,
[arXiv:hep-th/0308074].

\bibitem{Alday} J. A. Minahan, K. Zarembo, \textit{JHEP}\textbf{\ 0303}
(2003) 013, hep-th/0212208.

\bibitem{c1} A.V. Belitsky, S.E. Derkachov, G.P. Korchemsky, A.N. Manashov,
\textquotedblleft Superconformal operators in $N=4$ super-Yang-Mills
theory\textquotedblright , hep-th/0311104.

\bibitem{c2} Martin Kruczenski, \textquotedblleft Spin chains and string
theory\textquotedblright , hep-th/0311203.

\bibitem{t1} A.A. Tseytlin, \textit{Nucl. Phys.} \textbf{B664} (2003)
247-275, hep-th/0304139.

\bibitem{t2} S. Frolov, A.A. Tseytlin, \textit{Nucl.Phys.} \textbf{B668}
(2003) 77-110, hep-th/0304255.

\bibitem{t3} G. Arutyunov, S. Frolov, J. Russo, A.A. Tseytlin, \textit{Nucl.
Phys.} \textbf{B671} (2003) 3-50, hep-th/0307191.

\bibitem{t4} G. Arutyunov, J. Russo, A.A. Tseytlin, \textit{Phys. Rev.}
\textbf{D69} (2004) 086009, hep-th/0311004.

\bibitem{t5} M. Kruczenski, A.V. Ryzhov, A.A. Tseytlin, \textquotedblleft
Large spin limit of $AdS_{5}\times S^{5}$ string theory and low energy
expansion of ferromagnetic spin chains\textquotedblright , hep-th/0403120.

\bibitem{t6} M. Kruczenski, A. Tseytlin, \textquotedblleft Semiclassical
relativistic strings in $S^{5}$ and long coherent operators in $N=4$ SYM
theory\textquotedblright , hep-th/0406189.

\bibitem{w1} Bin Chen, Xiao-Jun Wang, Yong-Shi Wu, \textit{JHEP }\textbf{0402%
} (2004) 029, hep-th/0401016.

\bibitem{w2} Xiao-Jun Wang, Yong-Shi Wu, \textit{Nucl. Phys.} \textbf{B683}
(2004) 363-386, hep-th/0311073.

\bibitem{w3} Bin Chen, Xiao-Jun Wang, Yong-Shi Wu, \textit{Phys. Lett.}
\textbf{B591} (2004) 170-180, hep-th/0403004.

\bibitem{bpr} I. Bena, J. Polchinski, R. Roiban, \textit{Phys. Rev.} \textbf{%
D69} (2004) 046002, [arXiv:hep-th/0305116].

\bibitem{GS} M. B. Green, J. H. Shwarz, \textit{Phys. Lett.} \textbf{B 136}
(1984) 367-370, \textit{Nucl. Phys.} \textbf{B243} (1984) 285.

\bibitem{DNW} L. Dolan, C.R. Nappi, E. Witten, "A Relation Between
Approaches to Integrability in Superconformal Yang-Mills Theory", \textit{%
JHEP} \textbf{0310} (2003) 017, [arXiv:hep-th/0308089]; "Yangian Symmetry in
D=4 Superconformal Yang-Mills Theory", arXiv:hep-th/0401243.

\bibitem{bernard} D. Bernard, \textit{Commun. Math. Phys.} \textbf{137}
(1991) 191.

\bibitem{berkovits0001035} N. Berkevits, \textit{JHEP} 0004 (2000) 018,
arXiv:hep-th/0001035.

\bibitem{bracket} Faddeev, \textit{in \textquotedblleft Recent Advances in
Field Theory and Statistical Mechanics\textquotedblright , ed. J. Zabu and
R. Stora, Elsevier Sci. Pub. Amsterdam} 1984.

\bibitem{crmatrix} A. A. Belavin and V. G. Drinfeld, \textit{Soriet
Scientific Reviews c4} (1984), 93-165, \textit{Harwood Acad. Pub. New York.}

\bibitem{MT} R. R. Metsaev, A. A. Tseytlin, \textit{Nucl. Phys. }\textbf{B
533} (1998) 109-126, [arXiv:hep-th/9805028].

\bibitem{Met} R. R. Metsaev, \textit{JHEP} \textbf{0011} (2000) 014.

\bibitem{kallosh9805217} R. Kallosh, J. Rahmfeld, A. Rajaraman, \textit{JHEP}
\textbf{9809} (1998) 002, hep-th/9805217.

\bibitem{pesando9808020} I. Pesando, \textit{JHEP} \textbf{9811} (1998) 002,
hep-th/9808020.

\bibitem{b244} O. Babelon, L. Bonora, \textit{Phys. Lett.} \textbf{B 244}
(1990) 220-226.

\bibitem{RS} R. Roiban and W. Siegel, \textit{JHEP} \textbf{0011}(2000) 024.

\bibitem{killing} R. Kallosh, "Superconformal Actions in Killing Gauge",
arXiv:hep-th/9807206.

\bibitem{lu9805151} H. Lu, C. N. Pope, J. Rahmfeld, \textit{J. Math. Phys. }%
\textbf{40} (1999) 4518-4526, hep-th/9805151.

\bibitem{kallosh9805041} R. Kallosh, A. Rajaraman, \textit{Phys. Rev. }%
\textbf{D58} (1998) 125003, hep-th/9805041.

\bibitem{hou} B.Yu. Hou, B.Yuan. Hou and P. Wang, \textit{J. Phys. A: Math.
Gen} \textbf{18} (1985) 165-185.

\bibitem{hhbook} B.Yu. Hou, B.Yuan. Hou, \textquotedblleft \textit{%
Differential Geometry for Physicists\textquotedblright }, \textit{World
Scientific Publishing Company} (April 1, 1997).

\bibitem{Babelon} O. Babelon and D. Bernard, \textit{Int. J. Mod. Phys. A}%
\textbf{\ 8} (1993) 507.

\bibitem{R1} V.G. Drinfeld, \textquotedblleft Quantum
Group\textquotedblright , \textit{Proceedings of the Intermational Congress
of Mathematicians, Berkeley }(1987) 798-820.

\bibitem{R2} M. Jimbo, \textit{Lett. Math. Phys.} \textbf{10} (1985) 63-69.

\bibitem{R3} L. D. Faddeev, N. Yu. Reshetikin and L .A. Takhtajan,
\textquotedblleft Quantization of Lie Groups and Lie
Algebras\textquotedblright , in \textit{Algebraic Analysis, eds. M.Kashiwara
and T.Kawai, Academic, Boston} (1989) 129-139.

\bibitem{bouyale} Boyu Hou, \textit{Yale University reprint YTP} 80-29 OCT.
1980, \textit{appeared in Commun. theor. phys. Vol. }\textbf{1} No.3 (1982);
B. Y. Hou, M. L. Ge and Y. S. Wu, \textit{Phys. Rev. }\textbf{D24} (1981)
2238.

\bibitem{houmath} Boyu Hou, \textit{J. Math. Phys.} \textbf{25} (1984) 2325.

\bibitem{u1} K. Ueno, \textquotedblleft Infinite dimensional Lie Algebra
acting on chiral fields and the Riemann Hilbert problem\textquotedblright ,
\textit{RIMS preprint} 374 (1981).

\bibitem{rh} V. E. Zakharov and A. V. Mikhailov, Sov. Phys. \textit{JETP}
\textbf{47} (1978) 1017.

\bibitem{kac} V. G. Kac, \textquotedblleft \textit{Infinite dimensional Lie
algebras}\textquotedblright , 3rd ed. \textit{Cambridge Univ. Press,
Cambridge, }1990.

\bibitem{mirror} E. Witten, \textquotedblleft Mirror Manifolds And
Topological Field Theory\textquotedblright , hep-th/9112056.

\bibitem{AA} A.Alekseev and S.Snatushvili, \textit{Commun. Math. Phys. }%
\textbf{133} (1990) 353.

\bibitem{houyang} Bo-yu Hou, Wen-li Yang, \textquotedblleft A $\hbar $%
-deformed Virasoro Algebra as Hidden Symmetry of the Restricted sine-Gordon
Model\textquotedblright , hep-th/9612235.

\bibitem{abz} Alexander B. Zamolodchikov and Alexey B. Zamolodchikov,
\textit{Annals of physics} \textbf{120} (1979) 253-291.

\bibitem{frenkel} E. Frenkel and N. Reshetikhin, \textit{Commun. Math. Phys.}
\textbf{178} (1996) 237-266.

\bibitem{khoroshkin} S. Khoroshkin, D. Lebedev, S. Pakuliak,
\textquotedblleft Intertwining Operators for the Central Extension of the
Yangian Double\textquotedblright , \textit{prepint DFTUZ/95/28, ITEP-TH-15/95%
}, q-alg/9602030.

\bibitem{houzhao} Xiang-Mao Ding, Bo-Yu Hou, Liu Zhao, \textquotedblleft $%
\hbar $-(Yangian) Deformation of Miura Map and Virasoro Algebra\textit{%
\textquotedblright }, q-alg/9701014.

\bibitem{polykov} A. M. Polyakov, \textquotedblleft Conformal Fixed Points
of Unidentified Gauge Theories\textquotedblright , hep-th/0405106.
\end{thebibliography}
\end{document}